\documentclass[aps,twocolumn]{revtex4}

\usepackage{graphicx}
\usepackage{dcolumn}
\usepackage{bm}
\usepackage{setspace}

\begin{document}
\newcommand{\oneover}[1]{\frac{1}{#1}}
\newcommand{\partialD}[2]{\frac{\partial #1}{\partial #2}}
\newcommand{\partialDD}[2]{\frac{\partial^2 #1}{\partial #2^2}}
\newcommand{\clocktransition}{$^1$S$_0$ $\rightarrow$ $^3$P$_0$ }

\newcommand{\MTwoTwo}[4]{\left(\begin{matrix} #1 & #2 \\ #3 & #4\end{matrix}\right)}
\newcommand{\prefactor}{\frac{e^2}{m_e}}
\newcommand{\measuredIR}{4\pi\epsilon_0\times(1.6\pm0.5)\times10^{-31}$m$^3}

\newcommand\T{\rule{0pt}{2.6ex}}
\newcommand\B{\rule[-1.2ex]{0pt}{0pt}}

\title{Blackbody radiation shift of the $^{27}$Al$^+$ \clocktransition transition}
\thanks{Work supported by ONR and NIST}
\author{T. Rosenband}
\email{trosen@boulder.nist.gov}
\author{W. M. Itano}
\author{P. O. Schmidt}
\thanks{Present address: Institut f\"ur Experimentalphysik, Universit\"at Innsbruck, Austria}
\thanks{Supported by the Alexander-von-Humboldt Stiftung}
\author{D. B. Hume}
\author{J. C. J. Koelemeij}
\thanks{Present address: Institut f\"ur Experimentalphysik, Heinrich-Heine-Universit\"at D\"usseldorf, Germany}
\thanks{Supported by the Netherlands Organisation for Scientific Research (NWO)}
\author{J. C. Bergquist}
\author{D. J. Wineland}
\affiliation{National Institute of Standards and Technology, 325
Broadway, Boulder, CO 80305}

\begin{abstract}
The differential polarizability, due to near-infrared light at 1126
nm, of the $^{27}$Al$^+$ \clocktransition transition is measured to
be $\Delta\alpha=\measuredIR$, where
$\Delta\alpha=\alpha_P-\alpha_S$ is the difference between the
excited and ground state polarizabilities. This measurement is
combined with experimental oscillator strengths to extrapolate the
differential static polarizability of the clock transition as
$\Delta\alpha(0)=4\pi\epsilon_0\times(1.5\pm0.5)\times10^{-31}$m$^3$.
The resulting room temperature blackbody shift of
$\Delta\nu/\nu=-8(3)\times10^{-18}$ is the lowest known shift of all
atomic transitions under consideration for optical frequency
standards.  A method is presented to estimate the differential
static polarizability of an optical transition, from a differential
light shift measurement.
\end{abstract}

\maketitle

The blackbody radiation shift \cite{WMI1982BB} is a significant
shift in all room temperature atomic frequency standards, as can be
seen in Table \ref{BBshifts}. It ranges from
$|\Delta\nu/\nu|\approx2\times10^{-14}$ for $^{133}$Cs to
$|\Delta\nu/\nu|\approx8\times10^{-18}$ for $^{27}$Al$^+$, as
reported here. In order to reach a systematic uncertainty of
$|\Delta\nu/\nu|<10^{-18}$, the transitions with a large room
temperature blackbody shift may require a cryogenic operating
environment, while $^{27}$Al$^+$ merely requires knowledge of the
room temperature background with 5 K uncertainty.
\begin{table}
\caption{Room temperature blackbody shifts and uncertainties of
various species in use, or under consideration, as atomic frequency
standards. Where no uncertainty is given, it is unknown.  The
$^{199}$Hg$^+$ optical transition is not listed, because this
standard operates at 4.2 K, where the blackbody shift is reduced by
$10^7$ from the room temperature value.} \label{BBshifts}
\begin{tabular*}{0.45\textwidth}{@{\extracolsep{\fill}} l | c | c | l}
species & transition & $|\Delta\nu/\nu|\times10^{18}$ & reference\\
\hline
Al$^+$ \T \B & $^1$S$_0 \rightarrow ^3$P$_0$ & $8(3)$ & \emph{this work}\\
In$^+$ & $^1$S$_0 \rightarrow ^3$P$_0$ & $<70$ & \cite{Becker2001In115}\\
Ag     & $^2$S$_{1/2} \rightarrow ^2$D$_{5/2}$ & $190$ & \cite{Topcu2006}\\
Yb$^+$ & $^2$S$_{1/2} \rightarrow ^2$F$_{7/2}$ & $234(110)$ & \cite{Lea2006BB}\\
Hg     & $^1$S$_0 \rightarrow ^3$P$_0$ & $240$ & \cite{PalchikovHgBBR}\\
Mg     & $^1$S$_0 \rightarrow ^3$P$_0$ & $394(11)$ & \cite{Porsev2006BB}\\
Yb$^+$ & $^2$S$_{1/2} \rightarrow ^2$D$_{3/2}$ & $580(30)$ & \cite{Schneider2005Yb171}\\
Sr$^+$ & $^2$S$_{1/2} \rightarrow ^2$D$_{5/2}$ & $670(250)$ & \cite{Madej2004Sr88}\\
Ca     & $^1$S$_0 \rightarrow ^3$P$_1$ & $2210(50)$ & \cite{Sterr2004Ca40}\\
Yb     & $^1$S$_0 \rightarrow ^3$P$_0$ & $2400(250)$ & \cite{Porsev2006BB}\\
Sr     & $^1$S$_0 \rightarrow ^3$P$_0$ & $5500(70)$ & \cite{Porsev2006BB}\\
Cs     & F$=4 \rightarrow $F$=3$ & $21210(260)$ & \cite{Jefferts2005F1}\\

\end{tabular*}

\end{table}

We begin with a brief explanation of the blackbody shift, followed
by an estimate of the shift in $^{27}$Al$^+$ based only on published
oscillator strengths.  The uncertainty in this estimate motivated us
to measure the differential polarizability of the clock transition
due to near-infrared light.  This measurement allows a determination
of the blackbody shift with reduced uncertainty.

\subsection{Blackbody shift}
The blackbody shift results from off-resonant coupling of thermal
blackbody radiation to the two states comprising the clock
transition.  The scalar polarizability $\alpha_a$ of an atomic state
$a$ driven by an electric field at frequency $\omega$ is
\begin{equation}
\label{DefAlpha}
\alpha_a(\omega)=\prefactor\sum_i\frac{f_i}{\omega_i^2-\omega^2},
\end{equation}
with summation over all transitions connecting to state $a$ with
resonant frequency $\omega_i$, and oscillator strength $f_i$.  For
monochromatic radiation $E_0 \cos{\omega t}$, this polarizability
results in a dynamic Stark shift of $\Delta
E_a=-\frac{1}{4}E_0^2\alpha_a(\omega)$.  The clock transition
suffers a blackbody radiation shift of
\begin{equation}
\label{BBintegral} \Delta\nu = \frac{-1}{4\epsilon_0\pi^3
c^3}\int^\infty_{0}\Delta\alpha(\omega)
\frac{\omega^3}{e^{\hbar\omega/k_B T}-1} d\omega,
\end{equation}
where we integrate over the power spectral density of the blackbody
 electric field, and
$\Delta\alpha(\omega)=\alpha_{P}(\omega)-\alpha_{S}(\omega)$ is the
difference between excited and ground state polarizabilities.  Here
we first estimate the differential static polarizability
$\Delta\alpha(0)$. This result is used to estimate the differential
polarizability at blackbody frequencies $\Delta\alpha(\omega)$ for
$\omega \approx 2\pi c/(10$ $\mu\textrm{m})$.

\subsection{The case of $^{27}$Al$^+$}
The transitions from the $^1$S$_0$ and $^3$P$_0$ states that have
been included in our estimate are listed in Table \ref{Values1126}.
Oscillator strengths are taken from the \textit{NIST Atomic Spectra
Database} \cite{NISTAtomicSpectra} where available, and from the
\emph{Opacity Project} \cite{OpacityProject} otherwise. From this we
calculate for the $^3$P$_0$ state
$\alpha_{P}(0)=4\pi\epsilon_0\times3.65(73)\times10^{-30}$ m$^3$.
For the $^1$S$_0$ state
$\alpha_{S}(0)=4\pi\epsilon_0\times3.68(78)\times10^{-30}$ m$^3$.
Thus,
$\Delta\alpha(0)=4\pi\epsilon_0\times(-0.03\pm1.0)\times10^{-30}$
m$^3$. The room temperature blackbody spectrum ($E_{rms}=830$ V/m)
is centered at 10 $\mu$m wavelength.  This corresponds to a
frequency $\omega$ in Eq. (\ref{DefAlpha}), which is 50 times lower
than the lowest transition frequency $\omega_i$. We may use the
static polarizability without loss of accuracy, and find
$\Delta\nu=-\frac{1}{2h}\Delta\alpha(0) E_{rms}^2=(0.00\pm0.06)$ Hz,
or fractionally $\Delta\nu/\nu=(0\pm6)\times10^{-17}$, since
$\nu\approx1.1\times10^{15}$ Hz.  In order to operate Al$^+$ as a
frequency standard with fractional frequency uncertainty below
$6\times10^{-17}$, the blackbody shift must be calibrated
experimentally.

\subsection{Near-infrared Stark shift measurement}
Ideally, we would measure the shift of the clock transition due to a
known intensity of 10 $\mu$m radiation, since the room temperature
blackbody field is centered at this wavelength. However, the windows
of our experimental apparatus are opaque to wavelengths longer than
3 $\mu$m. Instead, we measure the Stark shift due to near-infrared
radiation, and use this measurement to estimate the blackbody shift.

\begin{figure} \centering
\includegraphics[width=3in]{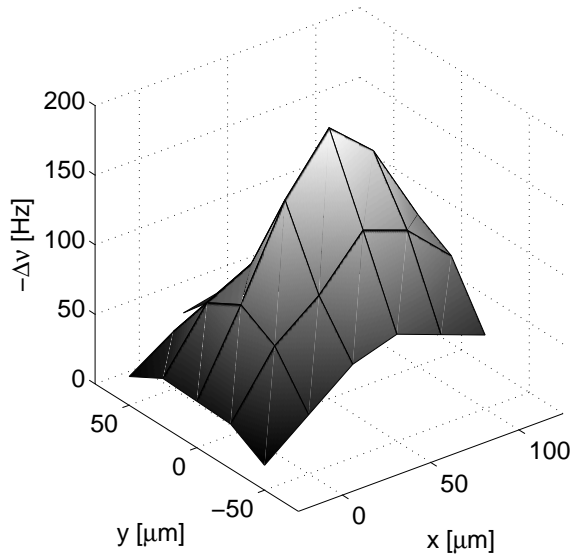}
\caption{Clock transition Stark shift vs. beam position. A Gaussian
beam profile fit yields $w_0=100$ $\mu$m for the beam waist, with a
peak shift of -190 Hz.} \label{measurement}
\end{figure}

The output of a fiber laser ($600$ mW with $\pm200$ mW fluctuations)
at 1126 nm was focussed onto an Al$^+$ ion, and switched on and off
at regular intervals. A stable ULE reference cavity was
simultaneously locked to the \clocktransition transition, via an
acousto-optic frequency shifter, and the frequency shift due to the
Stark shifting beam was tracked and recorded. These measurements
were repeated for various lateral (x,y) displacements of the Stark
shifting beam, as shown in Figure \ref{measurement}, in order to
estimate the beam waist ($w_0=100\pm10$ $\mu$m).

The resulting differential polarizability is $\Delta\alpha(2\pi
c/(1126$ nm$))=\measuredIR$, limited in accuracy by power
fluctuations of the Stark shifting laser.

\subsection{Extrapolation to zero frequency}
\label{ExtrapDC} The following relates this measurement to the
differential polarizability at 0 Hz, by expanding Eq.
(\ref{DefAlpha}) in small parameters. Two facts specific to Al$^+$
are used.
\begin{enumerate}
\item All strong transitions connecting to either clock state are in the deep
UV ($\lambda < 186$ nm).
\item The strongest transitions contributing to the sum in Eq. (\ref{DefAlpha}) are near each
other ($\lambda \approx 170$ nm).
\end{enumerate}

 Let $\delta_i \equiv (\omega/\omega_i)^2$. For the $^1$S$_0$ and
$^3$P$_0$ states in Al$^+$, $\delta_i < 0.03$, when $\omega = 2\pi
c/(1126 $ nm$)$. Expanding Eq. (\ref{DefAlpha}) in powers of
$\delta_i$ yields
\begin{equation}
\alpha(\omega) =
\alpha(0)+\prefactor\sum_i\frac{f_i}{\omega_i^2}(\delta_i +
\delta_i^2+...).
\end{equation}

Thus, the differential polarizability between the $^1$S$_0$ and
$^3$P$_0$ states is
\begin{equation}
\Delta\alpha(\omega)=\Delta\alpha(0)+
\prefactor\sum_i\frac{f_i}{\omega_i^2}(\delta_i+\delta_i^2+...),
\end{equation}
where we sum over all transitions connecting to the $^1$S$_0$ and
$^3$P$_0$ states.  Positive oscillator strengths are used for the
transitions connecting to $^3$P$_0$, and negative oscillator
strengths are used for the $^1$S$_0$ transitions.

Now let $\delta_0 \equiv (\omega/\omega_0)^2$, where
$\omega_0=2\pi$c/(171 nm), and let $\epsilon_i \equiv
\delta_i-\delta_0$. This value of $\delta_0$ is chosen because the
strong transitions all lie near 171 nm.  Then
\begin{equation}
\label{DefAlphaApprox} \Delta\alpha(0) = \frac{\Delta\alpha(\omega)-
\prefactor\sum_i\frac{f_i}{\omega_i^2}(\epsilon_i+\delta_i^2+...)}{1+\delta_0}.
\end{equation}

All of the terms after the summation sign are small, as can be seen
in Table \ref{Values1126}. For the strongest transitions
$\epsilon_i$ is small, because all strong transitions are near 171
nm.  For the weaker transitions $f_i/\omega_i^2$ is small. To test
the merits of this estimate, we propagate the uncertainties
$\sigma_{f_i}$ (see Table \ref{Values1126}) in the various $f_i$ via
Eq. (\ref{DefAlphaApprox}), which results in an uncertainty in
$\Delta\alpha(0)$ of

\begin{equation}
\label{DefError} \sigma_{\Delta\alpha(0)}\\
=\frac{e^2/m_e}{1+\delta_0}\sqrt{ \sum_i \left[\frac{\sigma_{f_i}}
{\omega_i^2}(\epsilon_i+\delta_i^2+...)\right]^2}.
\end{equation}

Note that our choice of $\delta_0$ minimizes the uncertainty
$\sigma_{\Delta\alpha(0)}$. Numerically we find
$\sigma_{\Delta\alpha(0)} = 4\pi\epsilon_0\times1.7\times10^{-33}$
m$^3 \approx 0.01\times\Delta\alpha(2\pi c/(1126$ nm$))$. Thus,
$\Delta\alpha(0)$ can be deduced from our measurement of
$\Delta\alpha(\omega)$ at 1126 nm with an additional uncertainty of
1 \%.  Eq. (\ref{DefAlphaApprox}) yields
$\Delta\alpha(0)=4\pi\epsilon_0\times(1.5\pm0.5)\times10^{-31}$
m$^3$.

\subsection{Estimate of blackbody shift}
Since the frequency of blackbody radiation (centered at 10 $\mu$m
wavelength) is closer to 0 Hz than the frequency of the applied 1126
nm radiation, we expect to relate $\Delta\alpha(0)$ to
$\Delta\alpha(2\pi c/(10$ $\mu$m$))$ with even less uncertainty than
our estimate of $\Delta\alpha(0)$ from $\Delta\alpha(2\pi c/(1126$
nm$))$.  As before, we can propagate the errors $\sigma_{f_i}$
through the result.  The calculation follows from Section
\ref{ExtrapDC} and Eq. (\ref{BBintegral}), and we simply write the
room temperature result as
\begin{equation} \Delta\nu = -\frac{\pi k_B^4 T^4}{60\epsilon_0 \hbar^4
c^3} (\Delta\alpha(0) \times 1.00024),
\end{equation}
or numerically, $\Delta\nu=-0.008(3)$ Hz.

\T \B
\subsection{Conclusion}
We have measured the differential polarizability of the
$^{27}$Al$^+$ \clocktransition clock transition at 1126 nm.  We have
also found expressions relating the differential polarizabilities at
various drive frequencies, in which the effect of uncertainties in
the oscillator strengths is minimized. In particular,
$\Delta\alpha(0)$ is found from $\Delta\alpha(2\pi c/(1126$ nm$))$
with 1 \% added fractional uncertainty, while allowing conservative
uncertainties of \linebreak 20 \% or larger in the oscillator
strengths. From $\Delta\alpha(0)$ we calculate the blackbody shift
with negligible added uncertainty. The fractional room temperature
blackbody shift $\Delta\nu/\nu = (-8\pm3)\times10^{-18}$ is
substantially lower for the $^{27}$Al$^+$ \clocktransition
transition than for other atomic frequency standards currently under
development (see Table \ref{BBshifts}).  The uncertainty in this
value could be lowered by improving the power stability of the 1126
nm Stark shifting laser.

\rule{0pt}{2.6ex} \rule[-1.2ex]{0pt}{0pt}

 This work is a contribution of NIST, an agency of the U.S.
government, and is not subject to U.S. copyright.

\begin{table*}
\caption{Transition wavelengths ($\lambda_i$) and oscillator
strengths ($f_i$) used to estimate $\Delta\alpha(0)$. Transitions
are in descending order of $f_i$ magnitude.  Negative oscillator
strengths are used for transitions connecting to the $3s^2$
$^1$S$_0$ ground state.  Fractional uncertainties
$\sigma_{f_i}/|f_i|$ were taken from the NIST Atomic Spectra
Database \cite{NISTAtomicSpectra} where available, and doubled.
Where no uncertainty is available, a fractional uncertainty of 1 is
assumed. $\delta_i = (\omega/\omega_i)^2$ where $\omega = 2\pi
c/(1126$ nm$)$, $\omega_i = 2\pi c/\lambda_i$, and $\epsilon_i =
\delta_i - (171/1126)^2$. The sixth column lists the summands of Eq.
(\ref{DefAlphaApprox}) truncated after $\delta_i^3$.}
\label{Values1126}
\begin{tabular*}{1\textwidth}{@{\extracolsep{\fill}} r | r | r | r | r | r | r | r | c}
$f_i$ & $\sigma_{f_i}/|f_i|$ & $\lambda_i$ & $\delta_i$ & $\epsilon_i$ &$\frac{e^2}{m_e}\frac{f_i}{\omega_i^2}(\epsilon_i + \delta_i^2 + \delta_i^3)$& from & to & ref.\\
 & & [nm] & & & [$4\pi\epsilon_0\times$m$^3\times10^{-33}$] & & & \\
\hline
-1.830000 & 0.20 & 167.079 & 0.0220 & -0.0010 & 2.005625 & 3s2 1S0 & 3s3p  1P1 & \cite{NISTAtomicSpectra}\\
0.903000 & 0.20 & 171.944 & 0.0233 & 0.0003 & 1.546615 & 3s3p 3P0 & 3s3d  3D1 & \cite{NISTAtomicSpectra}\\
0.612000 & 0.50 & 176.198 & 0.0245 & 0.0014 & 2.762716 & 3s3p 3P0 & 3p2   3P1 & \cite{NISTAtomicSpectra}\\
0.129000 & 0.20 & 185.593 & 0.0272 & 0.0041 & 1.541754 & 3s3p 3P0 & 3s4s  3S1 & \cite{NISTAtomicSpectra}\\
0.059000 & 1.00 & 118.919 & 0.0112 & -0.0119 & -0.701593 & 3s3p 3P0 & 3s4d  3D1 & \cite{NISTAtomicSpectra}\\
0.018000 & 1.00 & 104.789 & 0.0087 & -0.0144 & -0.202074 & 3s3p 3P0 & 3s5d  3D1 & \cite{NISTAtomicSpectra}\\
0.016556 & 1.00 & 120.919 & 0.0115 & -0.0115 & -0.196866 & 3s3p 3P0 & 3s5s  3S1 & \cite{OpacityProject}\\
0.005922 & 1.00 & 105.460 & 0.0088 & -0.0143 & -0.066807 & 3s3p 3P0 & 3s6s  3S1 & \cite{OpacityProject}\\
0.004078 & 1.00 & 98.598 & 0.0077 & -0.0154 & -0.043385 & 3s3p 3P0 & 3s6d  3D1 & \cite{OpacityProject}\\
-0.003020 & 1.00 & 93.527 & 0.0069 & -0.0162 & 0.030381 & 3s2 1S0 & 3s4p  1P1 & \cite{OpacityProject}\\
0.002889 & 1.00 & 98.905 & 0.0077 & -0.0153 & -0.030830 & 3s3p 3P0 & 3s7s  3S1 & \cite{OpacityProject}\\
0.001889 & 1.00 & 95.263 & 0.0072 & -0.0159 & -0.019393 & 3s3p 3P0 & 3s7d  3D1 & \cite{OpacityProject}\\
0.001656 & 1.00 & 95.429 & 0.0072 & -0.0159 & -0.017030 & 3s3p 3P0 & 3s8s  3S1 & \cite{OpacityProject}\\
-0.001100 & 1.00 & 71.470 & 0.0040 & -0.0190 & 0.007625 & 3s2 1S0 & 3s7p  1P1 & \cite{OpacityProject}\\
-0.001090 & 1.00 & 74.118 & 0.0043 & -0.0187 & 0.007995 & 3s2 1S0 & 3s6p  1P1 & \cite{OpacityProject}\\
-0.001050 & 1.00 & 69.949 & 0.0039 & -0.0192 & 0.007035 & 3s2 1S0 & 3s8p  1P1 & \cite{OpacityProject}\\
0.001050 & 1.00 & 93.341 & 0.0069 & -0.0162 & -0.010539 & 3s3p 3P0 & 3s9s  3S1 & \cite{OpacityProject}\\
0.001019 & 1.00 & 93.241 & 0.0069 & -0.0162 & -0.010214 & 3s3p 3P0 & 3s8d  3D1 & \cite{OpacityProject}\\
-0.000998 & 1.00 & 68.994 & 0.0038 & -0.0193 & 0.006541 & 3s2 1S0 & 3s9p  1P1 & \cite{OpacityProject}\\
-0.000948 & 1.00 & 68.353 & 0.0037 & -0.0194 & 0.006120 & 3s2 1S0 & 3s10p 1P1 & \cite{OpacityProject}\\
-0.000858 & 1.00 & 67.902 & 0.0036 & -0.0194 & 0.005480 & 3s2 1S0 & 3s11p 1P1 & \cite{OpacityProject}\\
0.000708 & 1.00 & 91.980 & 0.0067 & -0.0164 & -0.006985 & 3s3p 3P0 & 3s10s 3S1 & \cite{OpacityProject}\\
0.000613 & 1.00 & 91.916 & 0.0067 & -0.0164 & -0.006048 & 3s3p 3P0 & 3s9d  3D1 & \cite{OpacityProject}\\
-0.000567 & 1.00 & 79.448 & 0.0050 & -0.0181 & 0.004612 & 3s2 1S0 & 3s5p  1P1 & \cite{OpacityProject}\\
0.000501 & 1.00 & 91.041 & 0.0065 & -0.0165 & -0.004885 & 3s3p 3P0 & 3s11s 3S1 & \cite{OpacityProject}\\
0.000398 & 1.00 & 90.997 & 0.0065 & -0.0165 & -0.003876 & 3s3p 3P0 & 3s10d 3D1 & \cite{OpacityProject}\\
0.000124 & 1.00 & 70.040 & 0.0039 & -0.0192 & -0.000835 & 3s3p 3P0 & 3p5p  3P1 & \cite{OpacityProject}\\
0.000116 & 1.00 & 65.800 & 0.0034 & -0.0196 & -0.000701 & 3s3p 3P0 & 3p6p  3P1 & \cite{OpacityProject}\\
0.000083 & 1.00 & 63.690 & 0.0032 & -0.0199 & -0.000477 & 3s3p 3P0 & 3p7p  3P1 & \cite{OpacityProject}\\
0.000059 & 1.00 & 62.480 & 0.0031 & -0.0200 & -0.000328 & 3s3p 3P0 & 3p8p  3P1 & \cite{OpacityProject}\\
0.000043 & 1.00 & 61.710 & 0.0030 & -0.0201 & -0.000233 & 3s3p 3P0 & 3p9p  3P1 & \cite{OpacityProject}\\
0.000032 & 1.00 & 61.190 & 0.0030 & -0.0201 & -0.000170 & 3s3p 3P0 & 3p10p 3P1 & \cite{OpacityProject}\\
0.000001 & 1.00 & 80.860 & 0.0052 & -0.0179 & -0.000009 & 3s3p 3P0 & 3p4p  3P1 & \cite{OpacityProject}\\
\hline\hline SUM         & & & & & 6.609223 & & & \\
UNCERTAINTY & & & & & 1.685575 & & & \\
\end{tabular*}
\end{table*}


\end{document}